\documentclass[RNAAS]{aastex62}

\graphicspath{{./}{figures/}}

\received{January 1, 2018}
\revised{January 7, 2018}
\accepted{\today}
\submitjournal{ApJ}

\shorttitle{Sample article}
\shortauthors{Holwerda et al.}

\begin{document}

\title{TESS as a Low Surface Brightness Observatory}

\correspondingauthor{Benne W. Holwerda}
\email{benne.holwerda@loisville.edu}

\author[0000-0002-4884-6756]{Benne W. Holwerda}
\affil{Department of Physics and Astronomy, 102 Natural Science Building, University of Louisville, Louisville KY 40292, USA}




\section{\label{s:intro}Introduction}

The low surface brightness (LSB, $\mu < 25$ mag/arcsec$^2$) Universe holds clues to the formation and evolution of galaxies. Two examples of this have attracted a substantial observational effort over the last few years: first the stellar halo of Milky Way type galaxies and substructures therein. The stellar halo holds the remnants of the precursors that the final galaxy was assembled from, a fossil record of the galaxy's formation \citep{Bullock05,Johnston08a}. Secondly, ultra-diffuse galaxies (UDG) have been reported \citep[e.g.][]{van-Dokkum15a,van-der-Burg17}. 

Nearly all the satellites of the Galaxy were discovered thus far through resolved star counts have extremely low surface brightness ($\sim$28 mag/arcsec$^2$). Fainter or more distant populations would remain undetectable. Progress in these two fields can only be achieved by measuring ultra-low surface brightness levels. 

The reason is that current instrumentation is not optimized for the discovery of extended,
low surface brightness objects but faint resolved sources. For low surface brightness sensitivity fast optics are required i.e. low focal length over diameter (f/D). 

\section{\label{s:tess}TESS}

{\em TESS} \citep{Ricker15} will send back full images every 30 minutes. Combined, these will result in deep coverage at high Ecliptic Latitude in a single optical/infrared band, limited only by zodiacal light with an effectively 27 to 351 days of exposure time. It has a set list of stars to monitor \citep{Stassun17} with a large yield of new exoplanets as the primary science return \citep{Sullivan15}. 

{\em TESS} is optimized for red optical sensitivity with a single broad filter covering 600-1100 micron. {\em TESS} optics are a fast refractor, as opposed to {\em Kepler} reflecting optics.

\subsection{Telescope Performance for Low-Surface Brightness Observations}
\label{s:telescopes}

The sensitivity limit of telescopes is not just ruled by the speed of the optics (f/D) or their size (D). \cite{Valls-Gabaud17} gives the surface brightness limit of a telescope using this expression:
\begin{equation}
SB_{extended} = (1-\epsilon) \pi^2 \left({ D \over f}\right)^2 t_{exp} s^2_{pix} N_{pix} 10^{-0.4\mu}
\end{equation}
where, $\epsilon$ is flatfielding efficiency, $D$ is the diameter of the telescope, $f$ is the focal length, $s_{pix}$ is the pixel size on the sky, $N_{pix}$ is the number of pixels in an LSB object, $\mu$ is the on-sky surface brightness of the object, and $t_{exp}$ is the exposure time, the only variable for an observer. 
To optimize for LSB observations, one needs {\em fast optics} and a well-characterized, clean PSF. Both considerations count against telescopes using mirrors. 

The Dragonfly and Huntsman telescopes \citep{Abraham14} are successful and specifically optimized for LSB, using an array of telephoto lenses. Dragonfly observations revealed a large number of UDG \cite{van-Dokkum15a,Merritt14} and characterized stellar haloes around nearby galaxies \citep{van-Dokkum15b,Merritt16,Merritt16a}. The Dragonfly is considered the state-of-the-art for LSB and typically observed single targets for tens of hours each.

\begin{figure}[htbp]
\begin{center}
   \begin{minipage}[tl]{\linewidth}
	\includegraphics[width=\linewidth]{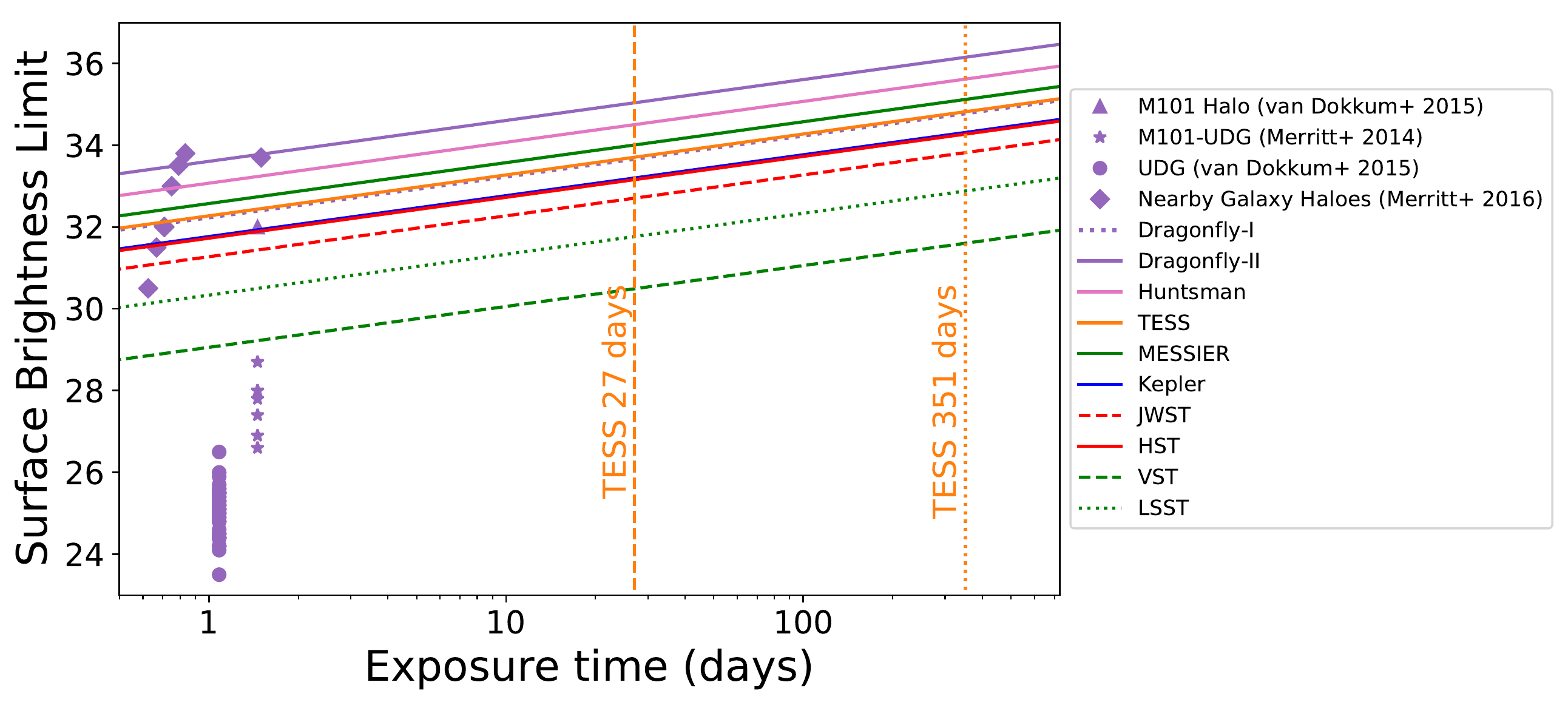}
	\caption{The approximate surface brightness limit as a function of exposure time for HST, JWST, VST, LSST,  Kepler, TESS, the MESSIER mission concept and the Dragonfly telescope (lines). Points are Dragonfly observations of stellar haloes \protect\citep{van-Dokkum15a,Merritt16a}, the ultra-diffuse galaxies nearby M101 \protect\citep{Merritt14}, and the ultra-diffuse galaxies in Virgo \protect\citep{van-Dokkum15b}. In principle, most telescopes can observe UDG when committing enough exposure time but stellar haloes will always remain challenging. Dragon-II and MESSIER are purposed to observe the LSB Universe but we note how well {\em TESS} performs within the existing observational schedule. }
\label{f:telescopes}
    \end{minipage}\hfill
\end{center}
\end{figure}

Figure \ref{f:telescopes} illustrates how all telescopes can eventually observe LSB features provided enough exposure time is committed. LSST and VST observations are sufficient for UDG identification \citep{van-der-Burg17}.
The Dragonfly, Huntsman, and {\em MESSIER} mission concept are purpose-built for the LSB universe but it noteworthy how close the TESS mission performs to these optimized observatories. 

{\em TESS} short/wide campaign is 27 days and the deep campaign at the poles is 351 days. Both are more than sufficient to observe LSB features such as stellar streams, stellar haloes and ultra-diffuse galaxies. 



The remaining issue is whether LSB features can be observed with enough contrast. This depends on how much background light (e.g. zodiacal) can be corrected for versus the expected peak luminosity of the features. \citep{Knapen17}

The UDG and stellar streams observed are all optically {\em red}. The observational drawback is that zodiacal and Galactic cirrus contribute at these wavelengths. 

The {\em TESS} campaign will reveal for how many nearby galaxies stellar halo properties and LSB companions can be observed.



\end{document}